\documentclass[10pt,epsf]{article}
 \usepackage{amssymb}
 \usepackage{amsthm}
 \usepackage{mathrsfs}
 \usepackage{pifont}
 \usepackage{amsmath}
 \usepackage{latexsym}
 \usepackage{graphicx}
  \newcommand{\integ}[2]{\displaystyle \int_{#1}^{#2}}

  \newcommand{\no}{\noindent}

 \newcommand{\ind}{1\!\!1}

 \newcommand{\findemo}{\hfill\mbox{\vrule height4pt width4pt depth0pt }}

 \newcommand{\beqar}{\begin{eqnarray}}
 \newcommand{\eeqar}{\end{eqnarray}}

 \def \cadlag {{c\`adl\`ag}~}
 \newtheorem{definition}{Definition}[section]
 \newtheorem{thm}{Theorem}[section]
 
 \newtheorem{lemma}{Lemma}[section]
 \newtheorem{remark}{Remark}[section]

 \newcommand{\be}{\begin{equation}}
\newcommand{\ee}{\end{equation}}
\newtheorem{axiom}{Definition}
  \def \R{\mathbb{R}}
   \def \Y{\widetilde{Y}}
    
 \def \P{\mathbb{P}}

 \def \E{\mathbb{E}}
\def \Ff{\mathbb{F}}

 \def\M{\mathcal{M}}
 \def \F{\mathcal{F}}
 \def \A{\mathcal{A}}

 \def \bf{\textbf}
 \def \it{\textit}
 \def \S{\mathcal{S}}
 \def \ds{\displaystyle}
 
\def \essup {\mbox{ess sup}}
\def \essinf {\mbox{ess inf}}

\begin{document}
\title{Optimal stopping of expected profit and cost yields in an investment under uncertainty}
\author{Boualem Djehiche,\thanks{Department of Mathematics, The Royal
Institute of Technology, S-100 44 Stockholm, Sweden.\@ e-mail:
boualem@math.kth.se}\,\,\,\, Said Hamad\`ene\thanks{Universit\'e du
Maine, D\'epartement de Math\'ematiques, Equipe Statistique et
Processus, Avenue Olivier Messiaen, 72085 Le Mans, Cedex 9, France.
e-mail: hamadene@univ-lemans.fr}\,\,\, and \, Marie-am\'elie
Morlais\thanks{Universit\'e du Maine, D\'epartement de
Math\'ematiques, Equipe Statistique et Processus, Avenue Olivier
Messiaen, 72085 Le Mans, Cedex 9, France. e-mail:
 Marie\_Amelie.Morlais@univ-lemans.fr}}

\date{\today}
\maketitle


\begin{abstract}
We consider a finite horizon optimal stopping problem related to
trade-off strategies between expected profit and cost cash-flows of
an investment under uncertainty. The optimal problem is first
formulated in terms of a system of  Snell envelopes for the profit
and cost yields which act as obstacles to each other. We then
construct both a minimal and a maximal solutions using an
approximation scheme of the associated system of reflected backward
SDEs. We also address the question of uniqueness of solutions of this system of SDEs. 
When the dependence of the cash-flows on the sources of uncertainty, such as fluctuation
market prices, assumed to evolve according to a diffusion process,
is made explicit, we also obtain a connection between these
solutions and viscosity solutions of a system of variational
inequalities (VI) with interconnected obstacles.
\end{abstract}
{\bf AMS Classification subjects}: 60G40 ; 93E20 ; 62P20 ; 91B99.

\medskip
\section{Introduction}

The trade-off between the expected profit and cost yields is a
central theme in the cash-flow analysis of any investment project or
any industry which produces a commodity or provides services that
are subject to uncertainties such as fluctuating market prices or
demand and supply flows (see \cite{DixitPindyck} and
\cite{Trigeorgis} and the references therein). The project is
profitable when the expected profit yield is larger than the
expected cost yield, a relationship that cannot always be sustained,
due to many sources of uncertainty. Timing exit from the project
based an optimal trade-off between expected profit and cost yields
is thus a crucial decision.

An approach to this problem, which is widely used in portfolio
choice with transaction costs (see \cite{Korn} and the references
therein) is to impose a predetermined form of the cost yield and
formulate an optimal stopping or impulse control problem for the
expected profit yield, to determine exit and re-entry strategies.
But, in many investment projects subject to uncertain demand and
supply flows, the expected cost yield cannot be fully captured with
a given predetermined model.

In this work we do not assume any predetermined model for the cost
yield. We rather approach the problem by formulating a finite
horizon optimal stopping problem that involves both the expected
profit and cost yields which will act as obstacles to each other.
More precisely, given the profit (resp. cost) $\psi_1(t) dt$ (resp.
$\psi_2(t) dt$) per unit time $dt$, and the cost $a(t)$ (resp. profit
$b(t)$) incurred when exiting/abandonning the project, if we let
$Y^1$ and $Y^2$ denote the expected profit and the cost yields
respectively, the decision to exit the project at time $t$ depends
on whether $Y^1_t\ge Y^2_t-a(t)$ or $Y^2_t\le Y^1_t+b(t)$. \noindent
If $\F_t$ denotes the history of the project up to time $t$, the
expected profit yield at time $t$, is expressed in terms of a Snell
envelope as follows:
\begin{equation}\label{Y1}
Y^1_t={\essup}_{\tau\ge t}\E\left[\int_t^{\tau}\psi_1(s)ds+(Y^2_{\tau}-a(\tau))\ind_{[\tau <T]} +\xi_1\ind_{[\tau
=T]}| \F_t\right],
\end{equation}
where the supremum is taken over all exit times $\tau$ from the
project.
 Moreover for any $t\leq T$, the random time
\begin{equation}\label{tau}
\tau^*_t= \inf\{s\ge t,\; Y^1_s=Y^2_s-a(s)\}\wedge T,
\end{equation}
related to the cost $Y^2-a$ incurred when exiting the project should
be an optimal time to abandon the project after $t$, in which case,
we should also get:
\begin{equation}\label{Y1*}
Y^1_t=\E\left[\int_t^{\tau^*_t}\psi_1(s)ds+(Y^2_{\tau^*_t}-a(\tau^*_t))\ind_{[\tau^*_t
<T]}+\xi_1\ind_{[\tau^*_t =T]} |\F_t\right].
\end{equation}
In a similar fashion, the expected cost yield at time $t$ reads
\begin{equation}\label{Y2}
Y^2_t={\essinf}_{\sigma\ge t}\E\left[\int_t^{\sigma}\psi_2(s)ds+(Y^1_{\sigma}+b(\sigma) )\ind_{[\sigma
<T]}+\xi_2\ind_{[\sigma =T]} |\F_t\right],
\end{equation}
where, the infimum is taken over all exit times $\sigma$ from the
project. The random time
\begin{equation}\label{sigma}
\sigma^*_t=\inf\{s\ge t,\;  Y^2_s=Y^1_s+b(s)\}\wedge T
\end{equation}
related to the profit $Y^1+b$ incurred when exiting the project
should be optimal after $t$ as well.  In this case, we should get:
\begin{equation}\label{Y2*}
Y^2_t=\E\left[\int_t^{\sigma^*_t}\psi_2(s)ds+(Y^1_{\sigma^*_t}+b(\sigma^*_t))\ind_{[\sigma^*_t
<T]}+\xi_2\ind_{[\sigma^*_t =T]} |\F_t\right].
\end{equation}

In other words, the cost $Y^2-a$ and the profit $Y^1+b$ act as
obstacles that define the exit strategy.

The main result of the paper is to show existence of the pair
$(Y^1,Y^2)$ that solves the system of equations (\ref{Y1}) and
(\ref{Y2}) and also to prove that $\tau^*$ and $\sigma^*$ given
respectively by (\ref{tau}) and (\ref{sigma}) are optimal strategies
for our problem. Using the relation between Snell envelopes,
reflected backward SDEs (RBSDEs) and variational inequalities (see
[3] for more details), it then follows that solving the system of
equations (\ref{Y1}) and (\ref{Y2}) is equivalent to finding a
solution to the following reflected backward SDEs with
interconnected obstacles: for all $t\leq T$,
\[ (\S)\qquad \left\{
\begin{array}{l}
 Y_{t}^{1} = \xi^1+ \ds{\int_{t}^{T}\psi_{1}(s)
ds +
(K_{T}^{1} - K_{t}^{1} ) - \int_{t}^{T} Z_{s}^{1}dB_{s}};\\
Y_{t}^{2}  = \xi^2+\ds{\int_{t}^{T}\psi_{2}(s) ds
-(K_{T}^{2}- K_{t}^{2})- \int_{t}^{T} Z_{s}^{2}  dB_{s}}; \\
Y_{t}^{1}\ \ge  Y_{t}^{2} - a(t)\mbox{ and }
 Y_{t}^{2} \le Y_{t}^{1}+ b(t);\\
 \int_0^T \left(Y_s^1-(Y_s^2-a(s))\right)dK^1_s=0 \mbox{ and }\int_0^T (Y_s^1+b(s)-Y_s^2)dK^2_s=0.
\end{array} \right.
\]
Using an approximation scheme for systems of reflected BSDEs we
establish existence of both a ma-ximal and minimal solution of
$(\S)$. When the dependence of the cash-flows $(Y^1,Y^2)$ on the
sources of uncertainty, such as fluctuation market prices, that are
assumed to evolve according to a diffusion process $X$, is made
explicit, we also obtain a connection between the solutions of the
system $(\S)$ and viscosity solutions of the following system of
variational inequalities with interconnected obstacles:
\[  (VI)\;\;\; \left\{
\begin{array}{l}
\displaystyle{\min \{u^{1}(t, x)-u^{2}(t, x)+a(t,x), -\partial_{t}
u^{1}(t, x) - \mathcal{L}u^{1}(t, x)- \psi_{1}(t,
x ) \} =0},\\
\displaystyle{\min \{ u^{1}(t, x)+ b(t,x)- u^{2}(t, x),-\partial_{t}
u^{2}(t, x) - \mathcal{L}u^{2}(t, x)- \psi_{2}(t, x) \} } 0,\\u^1(T,x)=g_1(x),\,\,u^2(T,x)=g_2(x).
\end{array} \right.
\]

\medskip
The paper is organized as follows: Section 2 is devoted to the
formulation of the optimal stopping problem under consideration. In
Section 3, we construct a minimal and a maximal solution of $(\S)$,
using an approximation scheme, where the minimal solution is
obtained as a limit of an increasing sequence of solutions of a
system of reflected BSDEs, while the maximal one is obtained as a
limit of a decreasing sequence of solutions of another system of
reflected BSDEs. Next we address the question of uniqueness of the
solution of ($\cal S$). In general, uniqueness does not hold as it is shown through two
counter-examples. However, we give some sufficient conditions on $\psi^1, \psi^2, a$ and $b$,
for which a uniqueness result is derived. Finally, in Section 4, we establish a connection between
the solutions of the system $(\S)$ and viscosity solutions of the
system of variational inequalities with interconnected obstacles
$(VI)$. We actually show that $(VI)$ admits a solution. Uniqueness and finer regularity properties of the solutions of $(VI)$
require heavy PDE techniques which we prefere not include in this paper and will appear elsewhere.

\section{Preliminaries and the main result}
In this section we introduce some basic notions and results
concerning reflected BDSEs, which will be needed in the subsequent
sections.

Throughout this paper, $T>0$ denotes an arbitrarily fixed time
horizon, and $(\Omega, \F , \P)$ is a given probability space on
which is defined a $d$-dimensional Brownian motion $B
=(B_{t})_{0\le t\leq T}$. We also denote by $\Ff=(\F_t)_{0\le t\le T}$
the filtration generated by $B$ and completed by the $\P$-null sets
of $\F$. Throughout the sequel, we always denote by $B$ the process
restricted to $[0, T]$ and assume that all processes are defined on
$[0, T]$.

We shall also introduce the following spaces of processes which will
be frequently used in the sequel:
\begin{itemize}
 \item $\S^{2}$ is the set of all continuous $\Ff$-adapted processes $Y = (Y_{t})$
 such that $\E[\displaystyle{sup_{t \in [0, T]}|Y_{t}|^{2}}]< \infty$ 
\item $\A^{2}$ is the subset of $\S^2$ of increasing processes $(K_t)_{t\leq T}$ with $K_0=0$ ;
     \item $\M^{d, 2}$ denotes the set of $\Ff$-adapted and $d$-dimensional processes $Z$ such that
    $\E\left(\int_{0}^{T} |Z_{s}|^{2}ds \right)< \infty$.
\end{itemize}
The following results on reflected BSDEs are by now well known. For
a proof, the reader is referred to \cite{Elkarouietal97}. A solution
for the reflected BSDE associated with a triple ($f, \xi ,S$), where
$f: (t,\omega, y, z) \mapsto f(t,\omega,y, z)$ ($\R$-valued) is the
generator, $\xi$ is the terminal condition $\xi$ and $S:=(S_t)_{t\leq T}$ is
the lower barrier, is a triple $(Y_t, Z_t,K_t)_{0\le t\le T}$ of
$\Ff$-adapted stochastic processes that satisfies:
\begin{equation} \label{rbsde}\left\{
\begin{array}{ll}
Y\in \S^2,\,K\in \,\A^2 \mbox{ and }Z\in \M^{d,2}, \\Y_{t} = \xi+
\int_{t}^{T}f(s,\omega,Y_s,Z_s)ds + (K_{T} - K_{t}) - \int_{t}^{T}
Z_{s}dB_{s}  ,
\\ Y_t\ge S_t, \;\;\; 0\le t\le T,\\
 \int_{0}^{T} (S_t-Y_t)dK_t=0. \end{array} \right.
\end{equation}

The RBSDE($f, \xi ,S$) is said standard if the following conditions
are satisfied:
\begin{itemize}
\item[\bf{(A1)}] The generator $f$ is Lipschitz with respect to $(y,z)$ uniformly in $(t,\omega)$ ;

\item[\bf{(A2)}] The process $(f(t,\omega, 0,0, 0))_{0\le t\leq T}$ is ${\Ff}$-progressively measurable and $dt\otimes dP$-square integrable ;

\item[\bf{(A3)}] The random variable $\xi$ is in $L^{2}\left(\Omega,\F_{T},
\P\right)$;

\item [\bf{(A4)}] The barrier $S$ is continuous $\Ff$-adapted
and satisfies: $\E[\displaystyle{\sup_{0\le s\le T}|S_{s}^{+}|^{2}} ] < \infty$ and $S_{T}\le \xi$, $\P$-a.s.
\end{itemize}

\begin{thm}(see \cite{Elkarouietal97}) \label{RBSDE} Let the coefficients $(f,\xi, S)$ satisfy assumptions (A1)-(A4).
Then the RBSDE (\ref{rbsde}) associated with $(f,\xi, S)$ has a
unique $\Ff$-progressively measurable solution ($Y, Z, K$) which
belongs to $\S^{2}\times\M^{d, 2}\times \A^{ 2}$. Moreover the
process $Y$ enjoys the following representation property as a Snell
envelope: for all $t\leq T$,
\begin{equation} \label{snellenvelope}
Y^{1}_t=\essup_{\tau \geq t}E[\int_t^\tau f(s, Y_{s}^{1 },
Z_{s}^{1}) ds + S_\tau 1_{[\tau <T]}+\xi_11_{[\tau=T]}|\F_t].
\end{equation} 
\end{thm}

The proof of Theorem \ref{RBSDE} is related to the following, by
now standard, estimates and comparison results for RBSDEs. For the
proof see Proposition 3.5 and Theorem 4.1 in \cite{Elkarouietal97}.

\begin{lemma}\label{aprioriestimates}
Let $(Y, Z, K)$ be a solution of the RBSDE $(f, \xi, S)$. Then there
exists a constant $C$ depending only on the time horizon $T$ and on
the Lipschitz constant of $f$ such that:
\begin{equation}\label{RBSDEbound}
\E\left(\sup_{0\le t\leq T}|Y_t|^2+ \int_{0}^T|Z_{s}|^2ds +|K_{T}|^2
\right) \le C\E\left(\int_{0}^{T}|f(s, 0, 0)|^{2}ds + |\xi|^2 +
{\sup}_{0\le t\le T}|S_t^{+}|^{2}\right). 
\end{equation}
\end{lemma}

\begin{lemma} (Comparison of solutions)\label{standardcomparison}
Assume that $(Y, Z, K)$ and  $(Y^{'}, Z^{'}, K^{'})$ are solutions
of the reflected BSDEs associated with ($f, \xi, S$) and
$(f^{'},\xi^{'}, S^{'})$ respectively, where only one of the two
generators $f$ or $f'$ is assumed to be Lipschitz continuous. If
\begin{itemize}
\item $\xi \le \xi^{'}$, $\P$-a.s.,
\item $f(t, y, z) \le f^{'}(t, y, z), \;d\P \otimes dt$-a.s. and for all ($y, z$),
\item $\mathbb{P}$-a.s., for all $\, t\leq T$, $\,\,\, S_{t} \le S^{'}_{t}$,
\end{itemize}
then
\begin{equation}
\mathbb{P}\textrm{-a.s.}\,\,\quad\mbox{for all}\quad  t\leq T, \quad Y_{t} \le
Y_{t}^{'}.
\end{equation}
\end{lemma}

\medskip
The previous results are also valid for RBSDE with upper barriers.
Indeed, if $(Y, Z, K)$ solves the RBSDE associated with $(f,\xi, U)$
with upper barrier equal to $U$, then ($-Y, -Z, K$) solves the RBSDE
associated with $(\tilde{f}, -\xi, S)$ with parameters given by: $S-U$ and $\tilde{f}(s, y, z) = -f(s, -y, -z)$. 
\medskip

The first objective in this paper is to study existence of solutions
of the coupled system of RBSDEs $(\S)$. Let us introduce the
following assumptions:
\begin{itemize}
\item[(\bf{B1})] For each $i=1,2$, the mappings $(t,\omega,y,z)\mapsto \psi_i(t,\omega,y,z)$ are Lipschitz in $(y,z)$ uniformly in $(t,\omega)$ meaning that there exists
$C>0$ such that:
$$
|\psi_i(t,\omega,y,z)-\psi_i(t,\omega,y',z')|\leq C(|y-y'|+|z-z'|),
\quad \mbox{for all} \,\,\, t,y,z,y',z'.$$ Moreover the processes
$(\psi^i(t,0,0,0))_{0\le t\leq T}$ are $\Ff$-progressively measurable and
$dt\otimes dP$-square integrable ;
\item[(\bf{B2})] The obstacles $(a(t, \omega))_{0\le t\leq T}$ and $(b(t, \omega))_{0\le t\leq
T}$ belong to $\S^2$ ;

\item[(\bf{B3})] The random variables $\xi^1$ and $\xi^2$ are
${\cal F}_T$-measurable and square integrable. Moreover we assume
that P-$a.s.,\,\,\, \xi^1-\xi^2\geq \max\{-a(T),-b(T)\}$.
\end{itemize}
Throughout the sequel, we will also make use of either one of the two following assumptions:
\begin{itemize}
\item[\bf{(B4)}] The process $(b(t))_{0\le t\leq T}$ is of It\^o type,
i.e., for any $t\leq T$,
\begin{equation}
b(t) = b(0) + \ds{ \int_{0}^{t}U_{s}^{2}ds +\int_{0}^{t} V_{s}^{2}
dB_{s}},
\end{equation}
for some $\Ff$-progressively measurable processes $U^{2}$ and $V^{2}$
which are respectively $dt\otimes dP$, integrable and square
integrable. 
\item[(\bf{B4}$^{'}$)]
The process $(a(t))_{0\le t\leq T}$ is of It\^o type,
i.e., for any $t\leq T$,
\begin{equation}
a(t) = a(0) + \ds{ \int_{0}^{t}U_{s}^{1}ds +\int_{0}^{t} V_{s}^{1}
dB_{s}},
\end{equation}
for some $\Ff$-progressively measurable processes $U^{1}$ and $V^{1}$
which are respectively $dt\otimes dP$, integrable and square
integrable. 
\end{itemize}

\noindent
\begin{remark}
Assumption \bf{(B4)} is required to prove the continuity of the minimal solution, which is obtained by using an increasing approximation scheme, whereas Assumption
 (\bf{B4}$^{'}$) is required to get the continuity of the maximal solution.\\
\end{remark}

\noindent
Let us now make precise, on the one hand, the notion of a solution and,
on the other hand, the notions of minimal and maximal solutions of
the system $(\S)$.

\begin{definition}\label{defiS} A 6-uplet of processes $(Y^{1},\;Z^{1},\;K^{1},\;Y^{2},\;Z^{2},\;K^{2})$ is
called solution of the system $(\S)$  if the two triples
$(Y^{1},\;Z^{1},\;K^{1})$ and $(Y^{2},\;Z^{2},\;K^{2})$ belong to
${\S}^{2} \times {\M}^{d, 2}\times {\A}^{2}$ and if it satisfies
$(\S)$.

The process $(Y^{1},\;Z^{1},\;K^{1},\;Y^{2},\;Z^{2},\;K^{2})$ is a
minimal solution of the system $(\S)$ if it is a solution of $(\S)$
and if whenever another 6-uplet of processes $(\tilde Y^{1},\tilde
Z^{1},\tilde K^{1},\tilde Y^{2},\tilde Z^{2},\tilde K^{2})$ is
solution of $(\S)$ then
$$
 P-a.s.\,\,\,\mbox{for all}\,\,\, t\leq T, \quad \tilde Y_t^{1} \ge Y_t^{1} \,\, \mbox{ and }\,\, \tilde Y_t^{2} \ge Y_t^{2},
$$
whereas, it is a maximal solution $(\S)$ if
$$
 P-a.s.\,\,\, \mbox{for all}\,\,\, t\leq T, \quad \tilde Y_t^{1} \le Y_t^{1}\,\,\mbox{ and }\,\,
 \tilde Y_t^{2} \le Y_t^{2}. 
$$
\end{definition}

\medskip
The following theorems, related to existence of minimal repectively maximal solutions of $\S$,  are the main results of the paper. 

\begin{thm}\label{S} Assume that the data $(\psi_1,\psi_2,\xi^1,\xi^2,a,b)$ satisfy Assumptions (\bf{B1})-(\bf{B4}). Then the system $(\S)$ of
RBSDEs associated with $(\psi_1,\psi_2,\xi^1,\xi^2,a, b)$ admits a
minimal solution $(Y^1,Y^2, Z^1, Z^2, K^1, K^2)$.
\end{thm}

\begin{thm}\label{Smax} Suppose that the data $(\psi_1,\psi_2,\xi^1,\xi^2,a, b)$ satisfy Assumptions (\bf{B1})-(\bf{B3})
and suppose, in addition, Assumption (\bf{B4}$^{'}$) on the process
$(a(t))_{t\leq T}$. Then, the system $(\S)$ of RBSDEs associated
with $(\psi_1,\psi_2,\xi^1,\xi^2,a, b)$ admits a maximal solution
$(Y^1,Y^2, Z^1, Z^2, K^1, K^2)$.
\end{thm}

\noindent The proof of Theorem \ref{Smax} can be obtained from Theorem \ref{S} by
considering the minimal solution of the system associated with 
$(-\psi_1(t,\omega,-y,-z),-\psi^2(t,\omega,-y,-z),-\xi^1,-\xi^2,-a, -b)$.

\medskip\noindent
The next section is devoted to the proof of Theorem \ref{S}.

\section{Proof of Theorem \ref{S}}

\noindent \underline{\it{Step 1}}: Construction of the sequences and
properties. \medskip

\noindent We first introduce two increasing approximation
schemes $({Y}^{1, n},{Z}^{1, n}, {K}^{1, n})$ and
$({Y}^{2,n},{Z}^{2, n}, {K}^{2, n})$ that converge to the
minimal solution of $(\S)$. 

\noindent Consider the following BSDEs defined recursively, for any $n\geq 1$, by:
\begin{equation}\label{bsden0}\left\{\begin{array}{l}
(Y^{1, 0}, \; Z^{1, 0})\in \S^2\times \M^{d,2}\\
Y_{t}^{1, 0} = \xi^1+\int_{t}^{T}\psi_{1}(s,Y_{s}^{1, 0}, Z_{s}^{1,
0})ds -\int_{t}^{T} Z_{s}^{1, 0}dB_{s},\quad t\leq T;
\end{array}\right.
\end{equation}
and for $n\geq 0$ and any $t\leq T$,
\[ ({\S}_{n}) \;\;\left\{
\begin{array}{l}
 Y_{t}^{2, n+1}  =\xi^2+ \int_{t}^{T}\psi_{2}(s,Y_{s}^{2, n+1}, Z_{s}^{2, n+1})  ds - (K_{T}^{2, n+1} - K_{t}^{2, n+1} ) -
\int_{t}^{T} Z_{s}^{2, n+1}  dB_{s}, \\
 Y_{t}^{2, n+1} \le Y_{t}^{1, n}+ b(t), \\
Y_{t}^{1, n+1} = \xi^1+\int_{t}^{T}\psi_{1}(s, Y_{s}^{1, n+1 },
Z_{s}^{1, n+1 }) ds + (K_{T}^{1, n+1} - K_{t}^{1, n+1} ) -
\int_{t}^{T}
Z_{s}^{1, n+1 }dB_{s},\\
Y_{t}^{1, n+1} \ge  Y_{t}^{2, n+1} - a(t), \\
\int_0^T (Y_s^{1,n+1}-(Y_s^{2,n+1}-a(s))dK^{1,n+1}_s=0\mbox{ and }
\int_0^T (Y_{s}^{1,n}+ b(s)-Y_{s}^{2,n+1})dK^{2,n+1}_s=0.
\end{array} \right.
\]

In view of Assumptions (\bf{B1})-(\bf{B4}), it is easily shown by induction that for any $n\geq
1$, the triples $(Y^{1,n}, Z^{1, n}, K^{1, n})$ and $(Y^{2,n}, Z^{2,
n}, K^{2, n})$ are well defined and belong to the space ${\S}^{2}
\times {\M}^{d,2}\times {\A}^{ 2}$, since the pair of processes
$(Y^{1,0}, Z^{1, 0})$ solution of (\ref{bsden0}) exists.
Additionally, by the
comparison Lemma \ref{standardcomparison}, we have $\P$-a.s., for all
$t\leq T$, $Y^{1,0}_t\leq Y_t^{1,1}$, and, using once more an induction argument, we have for all $n\ge 0$,
$$
\P-a.s., \quad\mbox{for all}\quad t\leq T, \; \; Y_t^{1,n} \le Y_t^{1,
n+1} \;\;\;\textrm{and} \;\;\; Y_t^{2,n+1} \le Y_t^{2, n+2}.
$$

Next, let us consider the following standard BSDE:
$$\left\{\begin{array}{l}
\bar Y^2\in \S^2 \mbox{ and }\bar Z\in {\cal M}^{d,2}\\\bar
Y_{t}^{2} =\xi^2+ \int_{t}^{T}\psi_{2}(s,\bar Y_{s}^{2}, \bar
Z_{s}^{2})  ds - \int_{t}^{T} \bar Z_{s}^{2} dB_{s},\quad 
t\leq T.\end{array}\right.$$ 
The solution of this equation exists
(see e.g. \cite{Pardouxpeng1990}). Furthermore, since the
process $K^{2,n}$ is non-decreasing then using standard comparison
theorem for BSDE's (see e.g. \cite{Pardouxpeng1990}) we obtain:
\begin{equation}\label{eq:dominationsecondecomp}
\P-a.s.,\quad\mbox{for all}\quad t\leq T, \,\,\, \quad Y^{2,n}_t\leq \bar Y_t.
\end{equation}
 Finally, let
$(\tilde Y,\tilde Z,\tilde K)$ be the solution of the following
reflected BSDE associated with $(\psi_1, \xi_1,\bar Y)$, i.e., for
any $t\leq T$,
$$\left\{\begin{array}{l}(\tilde Y, \tilde Z, \tilde K)\in
\S^2\times \M^{d,2}\times \S^2\\\tilde Y_{t}^{1}= \xi^1+\int_{t}^{T}\psi_{1}(s, \tilde Y_{s}, \tilde Z_{s}) ds +
(\tilde K_{T} - \tilde K_{t} ) - \int_{t}^{T}
\tilde  Z_{s}dB_{s},\\
\tilde Y_{t} \ge  \bar Y_{t} - a(t), \\
\int_0^T (\tilde Y_s-(\bar Y_s-a(s))d\tilde
K_s=0.\end{array}\right.$$ Again thanks to the comparison Lemma
\ref{standardcomparison} we have
\begin{equation}\label{eq:dominationpremierecomp}
P-a.s.,\quad\mbox{for all}\quad t\leq T, \,\,\, \quad Y^{1,n}_t\leq \tilde Y_t.
\end{equation} 
Therefore, from (\ref{eq:dominationsecondecomp}) and (\ref{eq:dominationpremierecomp}) it follows that
\begin{equation}\label{eq:domination_solution}
E\left[\sup_{n\geq 0}\sup_{0\le t\leq
T}(|Y^{1,n}_t|+|Y^{2,n}_t|)^2\right]<\infty.  \end{equation}
Moreover, using the estimates given in Lemma \ref{aprioriestimates} for standard RBSDEs, there existens a real constant $C\geq 0$ such that for all $n\ge 0$,
\begin{equation}\label{eq:uniformcontrol_rbsde}
E[\int_0^T\{|Z^{1,n}_s|+|Z^{2,n}_s|\}^2ds]+E[(K^{1,n}_T)^2+(K^{2,n}_T)^2]\leq
C.\end{equation}

Let $Y^1$ and $Y^2$ be two optional processes defined, for all $t\le T$, by
$$
Y^1_t=\lim_{n\rightarrow \infty}Y^{1,n}_t \quad\mbox{and}\quad
Y^2_t=\lim_{n\rightarrow \infty}Y^{2,n}_t.$$
\medskip

 \noindent \underline{\it{Step 2}}: Existence of a solution for ($\S $).
 \medskip

\noindent Since the processes $b$ and $Y^{1,n}$ are of It\^o type,
then thanks to a result by El-Karoui \it{et al.} (\cite{Elkarouietal97}, Proposition 4.2., pp. 713),
the process $K^{2,n}$ is absolutely continuous w.r.t. $t$. Moreover, we
have, for all $t\le T$,
$$
dK^{2,n}_t\leq
1_{[Y^{2,n+1}_t=Y^{1,n}_t+b_t]}\{ \psi_{2}(t,Y_{t}^{1, n}+b_t,
Z_{t}^{1, n}+V^2_t) +U^2_t+\psi_{1}(t,Y_{t}^{1, n}, Z_{t}^{1,
n})\}^+dt.
$$
Hence, by (\bf{B1}) and (\bf{B4}), there exists a constant $C\geq 0$ such that, for all $n\geq 1$,
$$E[\int_0^T(\frac{dK^{2,n}_t}{dt})^2dt]\leq
C.$$ 

In view of this estimate together with (\ref{eq:uniformcontrol_rbsde}), there exists  a
subsequence along which both
 $((\frac{dK^{2,n}_t}{dt})_{0\le t\leq T})_{n\geq 1}$,
$((\psi_{2}(t,Y_{t}^{2, n+1}, Z_{t}^{2, n+1}))_{0\le t\leq T})_{n\geq 1}$
and $((Z_{t}^{2, n+1})_{0\le t\leq T})_{n\geq 1}$ converge weakly in their respective spaces
to the processes $(k^2_t)_{t\leq T}$,
$(\varphi_2(t))_{t\leq T}$ and $(Z^2_t)_{t\leq T}$ which also belong
to $\M^{1,2}$, $\M^{1,2}$ and $\M^{d, 2}$, repectively.

Next, for any $n\geq 0$ and any stopping time $\tau$ we have
$$
Y_{\tau }^{2, n+1}  =Y_{0}^{2, n+1}
-\int_{0}^{\tau}\psi_{2}(s,Y_{s}^{2, n+1}, Z_{s}^{2, n+1})  ds +
K_{\tau}^{2, n+1} + \int_{0}^{\tau} Z_{s}^{2, n+1} dB_{s}.$$ 
Taking the weak limits in each side and along this subsequence yields
$$
Y_{\tau}^{2}  =Y_{0}^{2} -\int_{0}^{\tau}\varphi_2(s)ds -
\int_0^\tau k^2_sds+ \int_{0}^{\tau} Z_{s}^{2} dB_{s},\,\quad \P-a.s.$$
Since the processes appearing in each side are optional, using the
Optional Section Theorem (see e.g. \cite{DellacherieMeyer75},
Chapter IV pp.220), it follows that
\begin{equation}\label{eq:weakdecomposition} \P-a.s., \,\,\forall t\leq T, \quad Y_{t}^{2}  =Y_{0}^{2}
-\int_{0}^{t}\varphi_2(s)ds - \int_0^t k^2_sds+ \int_{0}^{t}
Z_{s}^{2} dB_{s}.\end{equation} 
Therefore, the process $Y^2$ is continuous. Relying both on Dini's Theorem and on Lebesgue's dominated convergence one, we also get
that
$$
\lim_{n\rightarrow \infty}E[\sup_{t\leq T}|Y^{2,n}_t-Y^2_t|^2]=0.
$$

\medskip\noindent
We will now focus on the convergence of $(Y^{1,n})_{n\ge 0}$. Using estimates
(\ref{eq:domination_solution}) and (\ref{eq:uniformcontrol_rbsde})
and applying then Peng's Monotone Limit Theorem (see \cite{Peng99})
 to the sequence ($Y^{1, n}, Z^{1, n}, K^{1,n}$)
we get that $Y^1$ is \cadlag. Moreover, there exist an $\Ff$-adapted
\cadlag non-decreasing process $K^1$ and a process $Z^1$ of
$\M^{d,2}$ such that $(Z^{1,n})_{n\geq 0}$ converges to $Z^1$ in
$L^p(dt\otimes dP)$ for any $p\in [1,2)$: Moreover, for any stopping time
$\tau$, the sequence $(K^{1,n}_\tau)_{n\geq 1}$ converges weakly to
$K^1_\tau$ in $L^2(\Omega, {\cal F}_\tau,dP)$. Relying now on the Snell envelope representation (see  \cite{Elkarouietal97}, Proposition 2.3, pp 705), we have,
 for any $n\geq 1$ and $t\leq T$, 
$$
Y^{1,n+1}_t=\essup_{\tau \geq t}E[\int_t^\tau \psi_{1}(s, Y_{s}^{1,
n+1 }, Z_{s}^{1, n+1 }) ds + (Y_{\tau}^{2, n+1} - a(\tau))1_{[\tau
<T]}+\xi_11_{[\tau=T]}|\F_t].
$$
As $(Y^{2,n})_{n\geq 1}$ converges in $\S^2$ to $Y^2$ then, by taking
the limit in each side of the previous equality, we get
$$
\P-a.s.,\quad\mbox{for all}\quad t\leq T,\; \; \,Y^1_t =\essup_{\tau \geq
t}E[\int_t^\tau \psi_{1}(s, Y_{s}^{1 }, Z_{s}^{1 }) ds +
(Y_{\tau}^{2} - a(\tau))1_{[\tau <T]}+\xi_11_{[\tau=T]}|\F_t],
$$
since $Y^1$ is \cadlag. This further implies that $Y^1$ is
continuous and, using Dini's theorem, that the convergence of
$(Y^{1,n})_{n\geq 1}$ to $Y^1$ holds in $\S^2$. Relying next on the
Doob-Meyer decomposition of the 'almost' supermartingale $Y^{1}$
(see also Theorem \ref{RBSDE}), there exist $Z^1$ and
$K^1$ such that, for all $t\leq T$, 
\[\left\{
\begin{array}{l}
 Y_{t}^{1} = \xi^1+ \ds{\int_{t}^{T}\psi_{1}(s,Y^1_s, Z^1_{s})
ds +
(K_{T}^{1} - K_{t}^{1} ) - \int_{t}^{T} Z_{s}^{1}dB_{s}};\\
Y_{t}^{1}\ \ge  Y_{t}^{2} - a(t)\mbox{ and }
  \int_0^T \left(Y_s^1-(Y_s^2-a(s))\right)dK^1_s=0.
\end{array} \right.
\]

Since the convergence of $(Y^{1,n})_{n\geq 1}$ to $Y^1$ holds in
$\S^2$, we can now rely on standard arguments and, in particular, on
It\^o's formula applied to $(Y^{2,n}-Y^{2,m})^2$ ($m,n \geq 0$) to
claim that $(Z^{2,n})_{n\geq 1}$ is a Cauchy sequence and therefore that it
converges to $Z^2$ in $\M^{d,2}$. Using this and taking into account
the decomposition obtained in (\ref{eq:weakdecomposition}) we
finally get, for any $t\leq T$,
\[ \left\{
\begin{array}{l}
 Y_{t}^{2} = \xi^2+ \ds{\int_{t}^{T}\psi_{2}(s,Y^2_s, Z^2_{s})
ds -\int_t^Tk^2_sds
-\int_{t}^{T} Z_{s}^{2}dB_{s}};\\
Y_{t}^{2} \le  Y_{t}^{1} +b(t).\end{array} \right.
\] Due to the weak convergence of $((\frac{dK^{2,n}_t}{dt})_{t\leq T})_{n\geq 1}$ to the process $ k^2$ and the strong
convergence of ($Y^{1,n}$) and $( Y^{2, n})$ in $\S^{2}$, it follows that
$$0=\int_0^T (Y_{s}^{1,n}+
b(s)-Y_{s}^{2,n+1})dK^{2,n+1}_s\rightarrow \int_0^T (Y_{s}^{1}+
b(s)-Y_{s}^{2})k^2_sds=0$$which implies that
$(Y^2,Z^2,K^2:=\int_0^.k^2_sds)$ is solution for the second part of
$(\S)$ and henceforth, the $6$-uplet $(Y^1,Z^1,K^1,Y^2,Z^2,K^2)$ is
a solution of $(\S)$.
\medskip

This solution is actually a minimal one. Indeed, if there is another
one $(\underbar Y^1,  \underbar Z^1,  \underbar K^1, \underbar Y^2,
\underbar Z^2, \underbar K^2)$ then, by comparison, we obviously
get $ \underbar Y^1\geq Y^{1,0}$ and then $ \underbar Y^2\geq
Y^{2,1}$. Finally by induction we have, for any $n\geq 1$, $
\underbar Y^1\geq Y^{1,n}$ and then $
 \underbar Y^2\geq Y^{2,n}$, which implies the
desired result after taking the limit as $n$ goes to $\infty$.
\findemo

\subsection{On the uniqueness of the solution of the system $(\cal S)$}
As we will show below in Secton 3.2, in
general, we do not have uniqueness of the solution of $(\S)$. However, in
some specific cases, such as in the following result, uniqueness holds.

\begin{thm}\label{uniquenessresult} Assume that

$(i)$ the mappings $\psi_1$ and $\psi_2$ do not depend on $(y,z)$,
i.e., $\psi_i:=(\psi_i(t,\omega))$, $i=1,2$

$(ii)$ the barriers $a$ and $b$ satisfy:
\begin{equation}\label{eq:nullset_condition}
\P-a.s.\,\,\,\int_0^T1_{[a(s)=b(s)]}ds=0.
\end{equation}
Then, the solution of $(\cal S$) is unique.
\end{thm}
\no $Proof$. The proof relies on the uniqueness of the solution of a
reflected BSDE with one lower barrier. Indeed, let $(Y^1,Y^2, Z^1,
Z^2, K^1, K^2)$ be a solution of $(\cal S)$ and, for $t\leq T$, let
us set $Y_t=Y^1_t-Y^2_t$, $Z_t=Y^1_t-Y^2_t$ and $K_t=K^1_t+K^2_t$.
Therefore, the triple $(Y,Z,K)$ belongs to ${\cal S}^2\times {\cal
M}^{d,2}\times {\cal A}^2$. Moreover, for any $t\leq T$, it satisfies \be
\label{bsdedif}\left\{
\begin{array}{l}Y_{t} =\xi_1-\xi_2+
\ds{\int_{t}^{T}\{\psi^{1}(u,\omega)-\psi^{2}(u,\omega)\} du +
(K_{T} - K_{t} ) - \int_{t}^{T} Z_{u}dB_{u}},
\\ Y_{t}\ \ge  \max\{ - a(t),- b(t)\}\mbox{ and }
\int_0^T \left(Y_s+\min\{a(s),b(s)\}\right)dK_s=0.\end{array}\right.
\ee 
Indeed, the two first relations being obvious, it only remains to
show the third one. But, 
\be\label{skocond}\begin{array}{ll}
\int_0^T \left(Y_{s}+\min\{a(s),b(s)\}\right)dK_s =&
\int_0^T \left(Y^1_s-Y_s^2+a(s)\right)1_{[a(s)\leq
b(s)]}d(K^{1}_s+K^{2}_s)\\&+\int_0^T\left(Y^1_s-Y_s^2+b(s)\right)1_{[a(s)>b(s)]}d(K^{1}_s+K^{2}_s).\end{array}
\ee 
However, 
$$\int_0^T \left(Y^1_s-Y_s^2+a(s)\right)1_{[a(s)\leq
b(s)]}dK^{1}_s=0,
$$ 
since, for any $t\leq T$,
$dK^1_t=1_{[Y^1_t-Y_t^2+a(t)=0]}dK^1_t$. On the other hand,
$$
\begin{array}{l} $$0\leq \int_0^T \left(Y^1_s-Y_s^2+a(s)\right)1_{[a(s)\leq
b(s)]}dK^{2}_s\leq \int_0^T
\left(Y^1_s-Y_s^2+b(s)\right)1_{[a(s)\leq b(s)]}dK^{2}_s=0,$$
\end{array}$$ since, for any $t\leq T$, it holds that
$dK^2_t=1_{[Y^2_t-Y_t^1+b(t)=0]}dK^2_t.$ 
In the same way, one can show that the second term in (\ref{skocond}) is null and therefore,
the third relation in (\ref{bsdedif}) holds true. It follows that
$(Y,Z,K)$ is a solution for the one lower barrier reflected BSDE
associated with $(\psi_1-\psi_2,\xi_1-\xi_2, \max\{-a(t),-b(t)\})$.
As the solution of this latter equation is unique by Theorem
\ref{RBSDE}, then for any solution $(Y^1,Y^2, Z^1, Z^2, K^1, K^2)$
of $(\S)$, the differences $Y^1-Y^2$ and $Z^1-Z^2$ and the
increasing process $K^1+K^2$ are unique.

\medskip\noindent
Next, let us express $K^1$ and $K^2$ in terms of $K$. For any
$t\leq T$, we claim that \be\label{k1fctk}\begin{array}{ll}
K^1_t&=\int_0^t1_{[Y^1_s-Y^2_s+a(s)=0]}dK^1_s\\
{}&=\int_0^t1_{[Y^1_s-Y^2_s+a(s)=0]}1_{[a(s)<b(s)]}dK^1_s+\int_0^t1_{[Y^1_s-Y^2_s+a(s)=0]}1_{[a(s)>b(s)]}dK^1_s,\end{array}\ee
where, to get this second equality, we make use of both the absolute continuity of $dK^{1}$, the increasing property of $K^{1}$
and the condition (ii) on the barriers to argue that
$$
\int_0^T1_{[Y^1_s-Y^2_s+a(s)]}1_{[a(s)=b(s)]}dK^1_s=\int_0^Tds1_{[Y^1_s-Y^2_s+a(s)]}1_{[a(s)=b(s)]}(\frac{dK^1_s}{ds})=0.$$
On the other hand,
$$
\int_0^T1_{[Y^1_s-Y^2_s+a(s)=0]}1_{[a(s)<b(s)]}dK^2_s=\int_0^T1_{[Y^1_s-Y^2_s+a(s)=0]}1_{[a(s)<b(s)]}1_{[Y^1_s-Y^2_s+b(s)=0]}dK^2_s=0.
$$
In a similar fashion, we have
$$
\int_0^T1_{[Y^1_s-Y^2_s+a(s)=0]}1_{[a(s)>b(s)]}dK^2_s=0.
$$
Therefore, going back to (\ref{k1fctk}) we obtain,
for all $t\leq T$, 
$$
K^1_t=\int_0^t1_{[Y^1_s-Y^2_s+a(s)=0]}dK_s,
$$ which
implies that $K^1$ is unique and then so is $K^2$. Writing next the
equations satisfied by $Y^1$ and $Y^2$ and taking the conditional
expectation (w.r.t. $\F_{t}$), we obtain their uniqueness. From
standard arguments, uniqueness of $Z^1$ and $Z^2$ follows
immediately. \findemo

\begin{remark}
\indent
This uniqueness result can be slightly generalized
to generators of the following forms:
$$
\psi_1(t,\omega,y,z)=\tilde \psi_1(t,\omega)+\alpha_t y +\beta_t z
\mbox{ and } \psi_2(t,\omega,y,z)=\tilde \psi_2(t,\omega)+\alpha_t y
+\beta_t z$$ where $\alpha$ and $\beta$ are $\P$-progressively
measurable bounded processes with values in $\R$ and $\R^d$
respectively. The proof is the same as the previous one, noting that
the process $Y= Y^{1}-Y^{2}$ solves a linear RBSDE, which is
explicitely solvable.
\end{remark}

\noindent For the sake of completeness, we now consider the more general case i.e. when
 condition (\ref{eq:nullset_condition}) on the barriers is no more satisfied.
 \medskip

\begin{lemma}\label{trivial_uniquenessresult}Under \bf {(B1)-(B3)} together with (\bf {B4}) or $(\bf
{B4}^{'}$), if the condition (\ref{eq:nullset_condition})  on the
barriers is no more satisfied then, on any interval $[\alpha,
\beta]$ of $[0,T]$, where $a \equiv b$, uniqueness of a solution for the system
($\S$) holds only in the trivial cases where neither the first
component $Y^1$ nor the second one $Y^2$ are reflected processes.
\end{lemma}

\no $Proof$: To prove this, let us consider a non trivial interval
$[\alpha, \beta]$ with $ 0<\alpha < \beta<T$ and where
$a(t)=b(t),\,\forall t\in [\alpha,\beta]$ and assume that uniqueness
of the solution of $(\cal S)$ holds. So let
$(\underline{Y}^1,\underline{Z}^1,\underline{K}^1,\underline{Y}^2,\underline{Z}^2,\underline{K}^2)$
be the minimal solution of $(\S)$ which is then equal to the maximal
one by uniqueness. Next let us consider the following system of
RBSDEs on the time interval $[\alpha,\beta]$: for any $s\in
[\alpha,\beta]$,
\[ (\S^{\textrm{min}})\; \;\left\{ \begin{array}{l}
    dY_{s}^{1} = -\psi^{1}(s, Y_{s}^{1}, Z_{s}^{1})ds + Z_{s}^{1}dB_{s};\,\,Y^1_\beta=\underline{Y}^1_\beta \\

   dY_{s}^{2} = -\psi^{2}(s, Y_{s}^{2}, Z_{s}^{2})ds + Z_{s}^{2}dB_{s}+ dK_{s}^{2};\,\,Y^2_\beta=\underline{Y}^2_\beta  \\
  Y_{s}^{2} \le Y_{s}^{1} +b(s), \; \; \int_\alpha^\beta \left(Y_{s}^{1} +b(s) -Y_{s}^{2}\right) dK_{s}^{2} =0.
    \end{array} \right.  \]
Note also that existence and uniqueness for $(\S^{\textrm{min}}) $ on
 $[\alpha, \beta]$ results from standard results for BSDEs (or RBSDEs with one barrier).\\
On the other hand, let us consider the following system of reflected
BSDEs which is similar to ($\S$) but on the time interval
$[0,\alpha]$ and which actually has a solution. For any $t\in
[0,\alpha]$,
\[\qquad \left\{
\begin{array}{l}
 \gamma_{t}^{1} = Y^1_{\alpha}+ {\int_{t}^{\alpha}\psi^{1}(s,\gamma^1_s, \theta^1_{s})
ds +
(\zeta_{\alpha}^{1} - \zeta_{t}^{1} ) - \int_{t}^{\alpha} \theta_{s}^{1}dB_{s}};\\
 \gamma_{t}^{2} = Y^2_{\alpha}+ {\int_{t}^{\alpha}\psi^{2}(s,\gamma^2_s, \theta^2_{s})
ds +
(\zeta_{\alpha}^{2} - \zeta_{t}^{2} ) - \int_{t}^{\alpha} \theta_{s}^{2}dB_{s}}; \\
 \gamma_{t}^{1}\ \ge   \gamma_{t}^{2} - a(t)\mbox{ and }
  \gamma_{t}^{2} \le  \gamma_{t}^{1}+ b(t);\\
 \int_0^\alpha \left( \gamma_s^1-( \gamma_s^2-a(s))\right)d\zeta^1_s=0 \mbox{ and }\int_0^\alpha ( \gamma_s^1+b(s)- \gamma_s^2)d\zeta^2_s=0,
\end{array} \right.
\]
where both $Y^1_{\alpha} $ (resp. $Y^2_{\alpha}$) stands for the value at time $t = \alpha$ of the first component $Y^1$ (resp.
 the second component $Y^2$) of the
unique solution to the system ($\S^{\textrm{min}}$) defined on $[\alpha, \beta]$.
Therefore the following process 
$$\begin{array}{l}
\{(\gamma^1_t,\theta^1_t,\zeta^1_t,\gamma^2_t,\theta^2_t,\zeta_t^2)\ind_{[t\leq
\alpha]}+(Y^1_t,Z^1_t,\zeta^1_\alpha,Y^2_t,Z^2_t,K^2_t-K^2_\alpha+\zeta_\alpha^2)\ind_{[\alpha<t\leq
\beta]}+ \\\qquad\qquad\qquad(\underline{Y}^1_t, \underline{Z}^1_t,
\underline{K}^1_t-\underline{K}^1_\beta+\zeta_\alpha^1,
\underline{Y}^2_t, \underline{Z}^2_t,
\underline{K}^2_t-\underline{K}^2_\beta+K^2_\beta-K^2_\alpha+\zeta_\alpha^2)\ind_{[\beta<t\leq
T]}\}_{t\leq T},\end{array}
$$
obtained by concatenation, is also a solution for $(\S)$. Using once more uniqueness for ($\S$) yields for
any $t\in [\alpha,\beta]$,
$$
\underline{Y}^1_t=Y^1_t \mbox{ and }\underline{Z}^1_t=Z^1_t.$$ It
implies that the process $\underline{Y}^1$ is not reflected on the
time interval $[\alpha,\beta]$. Now going back to the system
$(\S^{\textrm{min}})$ and making the reflection on $Y^2$ and not on
$Y^1$, we define a new system denoted by ($\S^{\textrm{max}}$) and therefore, proceeding as above, we obtain that the solution $(\bar{Y}^1, \bar{Y}^2)$
 on $[\alpha, \beta ]$ is such that the second component is not reflected. By uniqueness, we can claim that: $(\underline{Y}^1, \underline{Y}^2) \equiv (\bar{Y}^1, \bar{Y}^2) $ on $[\alpha, \beta ]$ and hence,
  $\underline{Y}^2$ is not reflected on $[\alpha,\beta]$. \findemo
\begin{remark}
On any time interval where the solution of ($\S^{\textrm{min}}$) provides a solution of the system ($\S$), it is straightforward
 to check, by using standard comparison result for BSDEs or RBSDEs (see Theorem \ref{standardcomparison}), that the solution ($\underline{Y}^1, \underline{Y}^2$) is the minimal solution of ($\S$).\\
Similarly, whenever the solution ($\bar{Y}^1, \bar{Y}^2$) associated to the system ($\S^{max}$) is a solution of ($\S$), it can be proved that it is the maximal one.
\end{remark}

\subsection{Non-uniqueness: two counter-examples}
In this section, we provide two explicit counter-examples of system
$(\S)$ where uniqueness does not hold.
\paragraph{First example.}
In this first example, we study the case when the two penalties $a$
and $b$ are equal to zero, and the terminal conditions $\xi^{1}$ and $\xi^{2}$ are
equal. In addition, the two generators $(s, \omega) \rightarrow \psi_{i}(s, \omega, y, z)$, for $i=1,\; 2$, are assumed
 to be independent of $(y,z)$ and satisfy:
\be\label{sgn} \P-a.s., \quad\mbox{for all}\;\;\; t\leq T, \quad
\psi_{1}(t,\omega)- \psi_2(t,\omega) < 0.\ee Therefore, this leads
to the following system of RBSDEs
\[ (\S_{1})  \left\{  \;
\begin{array}{l}
 Y_{t}^{1} =\xi^{1} +{\int_{t}^{T}\psi_{1}(s) ds -\int_{t}^{T}Z_{s}^{1}dB_{s} + \big(K_{T}^{1} -K_{t}^{1} \big)},\\
  Y_{t}^{2} =\xi^{1} +{\int_{t}^{T}\psi_{2}(s) ds -\int_{t}^{T}Z_{s}^{2}dB_{s} - \big(K_{T}^{2} -K_{t}^{2} \big)},\\
Y^1_t\ge Y^2_t, \quad t\le T, \\ \int_0^T \left(Y_{s}^{1} -Y_{s}^{2} \right)d(K_{s}^{1}+K_{s}^{2}) =0.\\
\end{array}\right. \]

Introduce $(Y^{1, \textrm{min}}, Y^{2,
\textrm{min}})$ (resp. $(Y^{1, \textrm{max}}, Y^{2, \textrm{max}})$)
as being the minimal (resp. the maximal) solution of the system
$(\S_{1})$ as constructed in Section 2.

Our objective is to establish that, even in this simple example,
uniqueness for the system ($\mathcal{S}_{1}$) fails to hold. To do
this, we prove that the minimal and maximal solutions for that
system do not coincide.

\noindent So let us consider an arbitrary solution $(Y^{1}, Y^{2})$ of the system $(\S_{1})$. If we set $\bar{Y} = Y^{1} -Y^{2}$ then we
 have, for all $t\le T$,
 \begin{equation}\label{eq:RBSDE} \left\{\begin{array}{l}\bar Y_t = \displaystyle{\int_{t}^{T}(\psi_{1}(s) - \psi_{2}(s))ds  +
 (\bar{K}_{T} -\bar{K}_{t}) + \int_{t}^{T}\bar{Z}_{s}dB_{s},
 }\\
\bar Y_t\geq 0, \quad t\le T, \\ \int_0^T\bar Y_td\bar K_t=0.
 \end{array}\right.
 \end{equation}
Hence, standard results imply that the solution of that
reflected BSDE is unique. Taking then conditional expectation w.r.t.
$\mathcal{F}_{t}$ in (\ref{eq:RBSDE}), it yields
 $$ Y_{t}^{1} -Y_{t}^{2} = \displaystyle{\mathbb{E}\left( \int_{t}^{T}(\psi_{1}(s) - \psi_{2}(s))ds  + (\bar{K}_{T} -\bar{K}_{t})  | \mathcal{F}_{t}\right) }.  $$
Therefore, thanks to both (\ref{sgn}) and the constraint condition on
$Y^{1} -Y^{2}$, we obtain the strict increasing property of
$\bar{K}$. Next for $i=1,2$ denoting by $K^{i,\textrm{min}}$ and by
$ K^{i, \textrm{max}}$ the pair of increasing processes associated
with each component of both the minimal and maximal solution, we are
going to show that
$$
dK_{t}^{1, \textrm{min}} \equiv 0\;\;\;\textrm{and}\;\;\; dK_t^{2,
\textrm{max}} \equiv 0,\,\,\,\mbox{ for any }\,\,\, t\le T.
$$
To this end, let us consider the following system: for any $s\le T$,
\be \label{equni}\left\{ \begin{array}{l}
    d\b Y_{s}^{1} = -\psi_{1}(s)ds +\b Z_{s}^{1}dB_{s};\,\,\,\b Y^1_T=\xi^1,\\
d\b Y_{s}^{2} = -\psi_{2}(s)ds +\b Z_{s}^{2}dB_{s}+ d\b K_{s}^{2};\,\,\,\b Y^2_T=\xi^1,  \\
  \b Y_{s}^{2} \le \b Y_{s}^{1} , \; \; \; \int_0^T \left(\b Y_{s}^{1} -\b Y_{s}^{2}\right) d\b K_{s}^{2}
  =0.
    \end{array} \right.\ee 
Then, for any $t\leq T$, we have
\[ \begin{array}{ll}
\b Y_{t}^{1} -Y_{t}^{1, \textrm{min}} & \; = \displaystyle{
-\int_{t}^{T}(\b Z_{s}^{1} -Z_{s}^{1, \textrm{min}})dB_{s} -\left( K_{T}^{1, \textrm{min}} - K_{t}^{1, \textrm{min}} \right)}.\\
\end{array}  \]
As $Y_{t}^{1} -Y_{t}^{1, \textrm{min}} \ge 0$, it follows from
standard arguments that $\;K_{t}^{1, \textrm{min}} = K_{T}^{1,
\textrm{min}}$ for any $t\le T$ and then $K^{1, \textrm{min}}\equiv
0$. Using now uniqueness of BSDEs, reflected or not, we obtain that
$\b Y^1\equiv Y^{1,\textrm{min}}$ and then $\b Y^2\equiv
Y^{2,\textrm{min}}$. In the same way, by considering the reflection on
the other equation in (\ref{equni}) we obtain that
$K^{2,\textrm{max}}\equiv 0$.\\
Next, using again the uniqueness of the triple $(\bar{Y}= Y^1 -Y^2,
\bar{Z} =  Z^1 -Z^2, \bar{K} =K^1 +K^2)$ solving the one lower barrier RBSDE (\ref{eq:RBSDE}),
it follows that
\[ \begin{array}{ll}
 \bar{K}  & = \; K^{1,\textrm{min}}+ K^{2,\textrm{min}} =K^{2,\textrm{min}}  \\
    & = \; K^{1, \textrm{max}}+ K^{2, \textrm{max}}=   K^{1, \textrm{max}}. \\
 \end{array} \]
The process $\bar{K}$ uniquely defined by (\ref{eq:RBSDE}) being
strictly increasing then, in view of the second line of the previous
equalities, $K^{1,\textrm{max}}$ and $K^{2,\textrm{min}}$ are also
strictly increasing. Consequently, both $Y^{1, \textrm{min}}$ and
$Y^{1, \textrm{max}}$ solve BSDEs with same generator $\psi^{1}$ and
same terminal condition $\xi^1$. However, $Y^{1, \textrm{min}}$
solves a standard BSDE without any reflection, whereas $Y^{1,
\textrm{max}}$ solves a reflected BSDE with the associated process $
K^{1,\textrm{max}}$ which is strictly increasing, which yields that
$ Y^{1, \textrm{min}} \neq Y^{1, \textrm{max}}$ and achieves the
proof. \findemo
\paragraph{Second example.} Let us
assume the following structure of the generators
$$
\psi_{1}(t,\omega,y)=y \; \;\textrm{and} \; \;
\psi_{2}(t,\omega,y)=2y. 
$$ We also assume that for any $t\leq T$,
$a_t=b_t=0$. Then, we are led to consider the following system, for all $t \in [0, T]$,
\[  (\mathcal{S}_{2}) \left\{  \quad
\begin{array}{l}
Y_{t}^{1} =1 +\int_{t}^{T} Y_{s}^{1}ds - \int_{t}^{T}Z_{s}^{1}dB_{s} + \big( K_{T}^{1} - K_{t}^{1}\big),\\
 Y_{t}^{1} \ge Y_{t}^{2}\; \textrm{and}  \; \int_{t}^{T} (Y_{s}^{1} -Y_{s}^{2})dK_{s}^{1}  =0,\\
Y_{t}^{2} = 1+ \int_{t}^{T} 2Y_{s}^{2}ds - \int_{t}^{T}Z_{s}^{2}dB_{s} -\big( K_{T}^{2} - K_{t}^{2}\big),\\
 Y_{t}^{1} \ge Y_{t}^{2}\; \textrm{and} \; \int_{t_1}^{T} (Y_{s}^{2} -Y_{s}^{1})d K_{s}^{2} =0 ,\\
\end{array} \right.
\]
\indent In  this second
example, we will show that the minimal (resp. the maximal) solution constructed via an
increasing (resp. a decreasing) scheme are not equal and therefore
uniqueness does not hold. We note that, considering
an arbitrary solution of the system ($\S_{2}$),
 the difference $Y^{1}-Y^{2}$ does not solve any more a RBSDE, which was the crucial 
fact we rely on in the previous example.

\medskip\noindent To prove that uniqueness does not hold for ($\S_{2}$), let us
consider the minimal (resp. maximal) solution ($Y^{1, \textrm{min}},
Y^{2, \textrm{min}}$) (resp. ($Y^{1, \textrm{max}},\; Y^{2,
\textrm{max}}$)) of ($\S_{2}$) which is given, for all $t\in
[0,T]$, by
\[\left\{  \quad
\begin{array}{l}
dY_{t}^{1, \textrm{min}} =- Y_{t}^{1, \textrm{min}}dt -
Z_{t}^{1,\textrm{min}}dB_{t}\quad\mbox{ and }\quad Y_{T}^{1, \textrm{min}}=1,
\\
dY_{t}^{2, \textrm{min}} = -2Y_{t}^{2, \textrm{min}}dt - Z_{t}^{2,
\textrm{min}}dB_{t}+dK_{t}^{2,
\textrm{min}}\quad\mbox{ and }\quad Y_{T}^{2, \textrm{min}}=1,\\
Y_{t}^{1, \textrm{min}} \ge Y_{t}^{2, \textrm{min}}\quad \textrm{and}
\quad \int_{0}^{T}(Y_{t}^{1, \textrm{min}} -
 Y_{t}^{2, \textrm{min}})dK_{t}^{2, \textrm{min}} =0,
\end{array} \right.
\]
and
\[\left\{  \quad
\begin{array}{l}
dY_{t}^{1, \textrm{max}} =- Y_{t}^{1, \textrm{max}}dt -
Z_{t}^{1,\textrm{max}}dB_{t}-dK_{t}^{1, \textrm{max}}\quad\mbox{ and }\quad
Y_{T}^{1, \textrm{max}}=1,
\\
dY_{t}^{2, \textrm{max}} = -2Y_{t}^{2, \textrm{max}}dt
- Z_{t}^{2, \textrm{max}}dB_{t}\quad\mbox{ and }\quad Y_{T}^{2, \textrm{max}}=1,\\
Y_{t}^{1, \textrm{max}} \ge Y_{t}^{2, \textrm{max}}\quad \textrm{and}
\quad \int_{0}^{T}(Y_{t}^{1, \textrm{max}} -
 Y_{t}^{2, \textrm{max}})dK_{t}^{1, \textrm{max}} =0,
\end{array} \right.
\]
meaning that $K^{1, \textrm{min}}\equiv 0$ and $K^{2,
\textrm{max}}\equiv 0$. But, the solution of the first system is
given, for all $t\leq T$, by 
$$
Y_{t}^{1, \textrm{min}}=Y_{t}^{2,
\textrm{min}}=e^{T-t}, Z_{t}^{1, \textrm{min}}= Z_{t}^{2,
\textrm{min}}=0 \mbox{ and }K_{t}^{2, \textrm{min}}=e^T(1-e^{-t}).$$
On the other hand, the one of the second system is given, for all $t\leq T$, by  
$$
Y_{t}^{1, \textrm{max}}=Y_{t}^{2,
\textrm{max}}=e^{2(T-t)}, Z_{t}^{1, \textrm{max}}= Z_{t}^{2,
\textrm{max}}=0 \mbox{ and }K_{t}^{1,
\textrm{max}}=\frac{1}{2}e^{2T}(1-e^{-2t}).$$ Therefore, as we can
see, uniqueness of ($\S_2$) does not hold in this case. Finally, let
us point out that in order to exhibit the solutions of the previous
systems we have kept in mind two facts: $(i)$ the solutions of those
systems are deterministic ; $(ii)$ we have used properties of the
Snell envelope of processes.  \findemo

\bf{Remark 3.3.}
As a slight generalization of the previous counter-example,
 let us consider here the more general case when the generators take the following
 form:
$$\psi^{1}(s, y) = \alpha_{s}^{1}y \;\mbox{ and } \;\psi^{2}(s, y) = \alpha_{s}^{2}y,  $$
where $\alpha^{1}$ and $ \alpha^{2}$ are deterministic functions,
integrable on $[0, T]$ and satisfy, $\P-a.s.$ $\alpha_{t}^{1} <
\alpha_{t}^{2}, \forall \; t\leq T$.

We are led to study the following system given, for all $t \in [0,
T]$, by
\[  (\mathcal{S}_{2}^{'}) \left\{  \quad
\begin{array}{l}
dY_{t}^{1} =-\alpha_{t}^{1}Y_{t}^{1}dt +Z_{t}^{1}dB_{t} -dK_{t}^{1},\,\,\, Y^1_T=1,\\
 dY_{t}^{2} =-\alpha_{t}^{2}Y_{t}^{2}dt + Z_{t}^{2}dB_{t}+dK_{t}^{2},\,\,\, Y^2_T=1,\\
 Y_{t}^{1} \ge Y_{t}^{2}\quad \textrm{and} \quad
 \int_0^T(Y_{t}^{2} -Y_{t}^{1})d K_{t}^{1} =\int_0^T(Y_{t}^{2} -Y_{t}^{1})d K_{t}^{2}=0. \\
\end{array} \right.
\]
Similarly as in the previous proof, we consider here the minimal
solution (resp. the maximal solution) of the previous system which
is denoted by ($Y^{1, \textrm{min}} , Y^{2, \textrm{min}} $) (resp.
by ($Y^{1, \textrm{max}}, Y^{2, \textrm{max}}$)). In addition and as
already mentioned in Remark 3.2, the increasing processes associated
with these solutions are such that:
$$K^{1, \textrm{min}} \equiv 0 \;\mbox{ and } \;K^{2, \textrm{max}}  \equiv 0. $$
Thus, $Y^{1, \textrm{min}} $ and $Y^{2, \textrm{max}} $ solve a standard BSDEs which, by uniqueness, implies that,
for all $t \in [0, T]$,
$$Y_{t}^{1, \textrm{min}} = \exp(\int_{t}^{T} \alpha_{s}^{1}ds),\,\,
Y_{t}^{2, \textrm{max}} = \exp(\int_{t}^{T}\alpha_{s}^{2}ds) \mbox{
and } Z_{t}^{1, \textrm{min}} =Z_{t}^{2, \textrm{max}} =0.$$ 
Next, since the data of the system are deterministic and using the
characterization of the Snell envelope of a process as the
smallest supermartingale which dominates it (see e.g.
\cite{Elka}) we get that
\begin{equation}\label{eq:expression_solminimale}
Y_{t}^{2, \textrm{min}} = Y_{t}^{1, \textrm{min}}, \;Z_{t}^{2,
\textrm{min}} =0 \;\mbox{ and } \;K_{t}^{2, \textrm{min}}
=\displaystyle{\int_{0}^{t}\left( \alpha_{s}^{2} -
\alpha_{s}^{1}\right) Y_{s}^{1, \textrm{min}}ds},\,\,\,   t \in
[0, T]. \end{equation} Thus, we have obtained the minimal solution of
$(\S_2^{'})$.

Concerning the maximal solution, we proceed in the same way and then we can
check that both $Y^{1, \textrm{max}} $, $Z^{1, \textrm{max}} $ and
$K^{1, \textrm{max}} $ are given by:
$$ Y_{t}^{1, \textrm{max}} =Y_{t}^{2, \textrm{max}},
\;Z_{t}^{1, \textrm{max}} =  0
  \;\mbox{ and } \; K_{t}^{1, \textrm{max}} = \displaystyle{\int_{0}^{t}\left( \alpha_{s}^{2} - \alpha_{s}^{1}\right) Y_{s}^{2, \textrm{max}}ds},\,\,\, t\in [0,T].$$

Therefore since $\alpha_1<\alpha_2$, we obtain, for $i= 1, 2$,
$Y^{i, \textrm{min}} < Y^{i, \textrm{max}}$, which entails that, in
this case, uniqueness does not hold for ($\mathcal{S}_{2}^{'} $) as
well. \findemo

\section{Study of the related system of variational inequalities}

\subsection{The system of variational inequalities}
In this section, we briefly describe the connection between
solutions of $(\mathcal{S})$ and existence results of viscosity
solutions of the following system of variational inequalities with
interconnected obstacles:
\[  (VI)\;\;\; \left\{
\begin{array}{l}
\displaystyle{\min \{u^{1}(t, x)-u^{2}(t, x)+ a(t, x), -\partial_{t}
u^{1}(t, x) - \mathcal{L}u^{1}(t, x)- \psi^{1}(t,
x) \} =0},\\
\displaystyle{\max \{ u^{2}(t, x)+ b(t, x)- u^{1}(t, x), -\partial_{t} u^{2}(t, x) - \mathcal{L}u^{2}(t, x)- \psi^{2}(t, x) \} } = 0,\\
u^1(T,x)=g^1(x) \mbox{ and }u^2(T,x)=g^2(x),
\end{array} \right.
\]
when the dependence of dynamics of the cash-flows $Y^1, Y^2$ of e.g.
the fluctuations of the market prices, $X$, assumed to be a
diffusion process, is made explicit. Note that the obstacles may depend on the diffusion process
$X$. In $(VI)$, $\mathcal{L}$ denotes the infinitesimal
generator of the diffusion $X$. 

\medskip For $(t,x)\in [0,T]\times \R^k$, let
$X^{t,x}:=(X^{t,x}_s)_{s\leq T}$ be the solution of the following
standard differential equation:
\begin{equation}\label{eq: forwardproc}
dX_{s}^{t, x} = \mu(s, X_{s}^{t, x} )ds + \sigma(s, X_{s}^{t, x})
dB_{s},\,\; T\ge s\ge t, \quad X_{s}^{t, x}=x,\,\,s\leq t,
\end{equation} where, the two functions $\mu := (\mu(t, x))$ and $\sigma
:= (\sigma(t, x))$ defined on $[0, T] \times \mathbb{R}^{k}$ and taking their respective
values in $\R^k$ and $\R^{k\times d}$, are uniformly
Lipschitz w.r.t. $x$ and have a linear growth. This means that, for
all $t$, $x$, $y$
\[ \text{(\bf{C1})}\;\;\left\{ \begin{array}{ll}
|\mu(t, x) - \mu(t, y)| + |\sigma(t, x) -\sigma(t, y)| \le C|x -y|,  \\
  |\mu(t, x)|+ |\sigma(t, x)| \le C\big(1+ |x|\big). \\
\end{array} \right. \]These properties ensure both existence and uniqueness of a solution for (\ref{eq: forwardproc}).
Additionally for any $\theta \geq 2$, there exists a constant $C$
such that for any $x\in \R^k$\be
 \E \left[ \sup_{0\le s\leq T} \left| X^{t,x}_s
\right|^\theta \right] \leq C(1+|x|^\theta).\label{estimx0} \ee

Next, in this setting, the infinitesimal generator $\mathcal{L}$ of
the diffusion $X:\widehat{=} X^{t,x}$, is defined, for any function
$\Phi$ in $\mathcal{C}^{1,2}\left( [0, T] \times \mathbb{R}\right)$,
as follows:
$$
\mathcal{L}\left(\Phi  \right)(t, x) = \langle{ \mu(t, x), \Phi (t,
x) \rangle} + \frac{1}{2}\textrm{Trace}\left( \sigma
\sigma^{T}D^{2}\Phi(t, x)\right),
$$
where $^T$ stands for the transpose operation. Let us now make the following
assumption on the functions $\psi_i$, $i=1,2$.
$$
(\bf{C2}) \left\{\begin{array}{l} \mbox{ The functions $\psi^{1}$
and $\psi^2$ do not depend on $(y,z)$, i.e., $\psi_1= \psi_{1}(t,
x)$ and $\psi_{2}= \psi_{2}(t, x)$, }\\ \mbox{they are jointly
continuous in $(t,x)$ and have polynomial growth, i.e., there exist
two positive}\\\mbox{ constants $q$ and $C$ such that, for all $(t,
x)$,}\mbox{$\label{croispoly}\,\,\, |\psi^{i}(t, x)| \le C(1+
|x|^{q}),\;\;\; i=1,2.$}\end{array}\right.
$$
\medskip

The assumption that each generator $\psi_{i}$ does not depend on
$(y,z)$ but only on $x$ is quite natural especially for applications
in economics. It means that the payoffs are not of recursive type,
and the utilities $\psi_1$ and $\psi_2$ depend only on the process
$X^{t,x}$ which stands e.g. for the price of a commodity such as the
electricity or oil price in the market.

\noindent Next, let $g_i:=(g_i(x))$, $i=1,2$, and $a:=(a(t,x))$, $b:=(b(t,x))$
be given functions defined respectively on $\R^k$ and $[0, T]
\times \mathbb{R}^{k}$, with values in $\R$ and satisfy

\[ \text{(\bf{C3})}\;\;\left\{ \begin{array}{l}
(i)\,\,\mbox{ $a$ and $b$ are of polynomial growth and belong to
$\mathcal{C}^{1,2}([0, T] \times \mathbb{R}^{d})$, and satisfy }\\
\quad\mbox{for all
$(t,x)\in [0,T]\times \R^k, \integ{0}{T}1_{[a(s,X^{t,x}_s)=b(s,X^{t,x}_s)]}ds=0$ };\\
(ii)\,\, \mbox{ $g^1$ and $g^2$ are continuous and of polynomial
growth; }
\\
(iii) \,\, g^1(x)-g^2(x)\geq \max\{-a(T,x),-b(T,x)\},\,\,\, x\in
\R^k.
\end{array}\right.\]
\medskip

With the help of the solutions of the system of reflected BSDEs $(\cal S)$,
we are going to show that $(VI)$ has a solution in viscosity sense
whose definition is the following

\begin{axiom}
Let $(u_1,u_2)$ be a pair of continuous functions on $[0,T]\times
\R^k$. It is called:

$(i)$ a viscosity supersolution (resp. subsolution) of the system
$(VI)$ if for any $(t_0,x_0)\in [0,T]\times \R^k$ and any pair of
functions $(\varphi_1,\varphi_2)\in (C^{1,2}([0,T]\times \R^k))^2$
such that $(\varphi_1,\varphi_2)(t_0,x_0)=(u_1,u_2)(t_0,x_0)$ and
for any $i=1,2$, $(t_0,x_0)$ is a maximum (resp. minimum) of
$\varphi_i -u_i$ then, we have
\be\label{defvisco}
\begin{array}{l}
\min\{u_1(t_0,x_0)-u_2(t_0,x_0)+a(t_0,x_0),-\partial_t\varphi_1(t_0,x_0)-{\cal
L}\varphi_1(t_0,x_0)-\psi_1(t_0,x_0)\}\geq 0 \,\,(resp. \leq 0),\\
\mbox{ and }\\
\max\{u_2(t_0,x_0)-u_1(t_0,x_0)-b(t_0,x_0),-\partial_t\varphi_2(t_0,x_0)-{\cal
L}\varphi_2(t_0,x_0)-\psi_2(t_0,x_0)\}\geq 0 \,\,(resp. \le
0).\end{array} \nonumber\ee

$(ii)$ a viscosity solution of the system $(VI)$ if it is both a
viscosity supersolution and subsolution. 
\end{axiom}

\subsection{Construction and regularity of viscosity solution of the system (VI)}
\indent Our objective is to construct and identify a continuous viscosity solution
for the system (VI) by relying both on standard results for the representation of viscosity solutions by
BSDEs (see e.g. \cite{Elkarouietal97}, Theorem 8.5) and on the results derived in the previous sections.\\
\indent For $(t,x)\in [0,T]\times \R^k$, let $(Y^{1,(t,x)},Y^{2,(t,x)},
Z^{1,(t,x)}, Z^{2,(t,x)}, K^{1,(t,x)}, K^{2,(t,x)})$ be a solution
of the following system of reflected BSDEs: for any $s\in [t,T]$,
\[(\tilde S)\,\, \left\{\begin{array}{lll} Y_{s}^{1, (t, x)}
=g^1(X_T^{(t,x)})+
\ds{\int_{s}^{T}\psi^{1}(u, X_{u}^{t,x}) du + (K_{T}^{1, (t, x)} - K_{s}^{1, (t, x)} ) - \int_{s}^{T} Z_{u}^{1, (t, x)}dB_{u}}, \\
Y_{s}^{2, (t, x)}  = g^2(X_T^{(t,x)})+ \ds{\int_{s}^{T}\psi^{2}(u,
X_{u}^{t,x})du -(K_{T}^{2, (t, x)}- K_{s}^{2, (t, x)})- \int_{s}^{T}
Z_{u}^{2, (t, x)}  dB_{u}}, \; \;
 \\ Y_{s}^{1, (t, x)}\ \ge  Y_{s}^{2, (t, x)} - a(s, X_{s}^{t,x})\,\,\, \mbox{ and }\,\,\, Y_{s}^{2, (t, x)}
 \le Y_{s}^{1, (t, x)}+ b(s, X_{s}^{t,x}),\\ \int_t^T \left(Y_s^{1, (t, x)}-
 (Y_s^{2, (t, x)}-a(s, X_{s}^{t,x}))\right)dK^{1, (t, x)}_s=0, \\
 \int_t^T (Y_s^{1,
 (t, x)}+b(s, X_{s}^{t,x})-Y_s^{2, (t, x)})dK^{2, (t, x)}_s=0.
\end{array} \right.
\]
Note that, thanks to Theorems 2.2, 3.1 and Assumptions (C1)-(C3),
especially the facts that the processes $(a(s,X^{t,x}_s))_{s\in
[t,T]}$ and $(b(s,X^{t,x}_s))_{s\in [t,T]}$ verify ($\textbf{B4}$)
and (\bf{B4}$^{'}$) respectively, both the minimal and maximal
 solution exist and furthermore, these two solutions coincide. Indeed, uniqueness holds since the functions $\psi_i$,
$i=1,2$, do not depend on $(y,z)$ and the barriers satisfy the
condition $(ii)$ of Theorem \ref{uniquenessresult}. Moreover this unique solution is obtained as a
limit of the increasing and decreasing schemes since the functions
$a:=(a(t,x))$ and $b:=(b(t,x))$ belong to ${\cal
C}^{1,2}([0,T]\times \R^k)$ and are of polynomial growth.
\medskip

\noindent According to the construction of the minimal solution of ($\tilde
S)$ we have, for any $s\in [t,T]$,
$$
(Y_s^{1,(t,x)},Y_s^{2,(t,x)})=\lim_{n\rightarrow
\infty}(Y_s^{1,(t,x);n},Y_s^{2,(t,x);n}),
$$ 
where,
$(Y^{1,(t,x);n},Y^{2,(t,x);n})$ are defined in the same way as in
($\S_{n}$) but with the specific data above. Thanks to Theorem 8.5 
in \cite{Elkarouietal97}, and by an induction argument there exist
deterministic functions $u^i_n:=(u^i_n(t,x))$, $i=1,2$, continuous
on $[0,T]\times \R^k$ such that: for any $s\in [t,T]$,
$$
(Y_s^{1,(t,x);n},Y_s^{2,(t,x);n})=(u^1_n(s,X^{t,x}_s),u^2_n(s,X^{t,x}_s)).
$$
Moreover there exist two positive constants $\alpha_1$ and
$\alpha_2$ such that, for any $(t,x)\in [0,T]\times \R^k$,
$$
|u^i_n(t,x)|+|u^i_n(t,x)|\le \alpha_1 (1+|x|^{\alpha_2}).$$ Finally,
the sequences $(u^i_n)_{n\geq 0}$, $i=1,2$, are increasing since
$Y^{i,(t,x);n}\le Y^{i,(t,x);n+1}$, $i=1,2$. Therefore, there exist
two deterministic lower semi-continuous functions $u^i$, $i=1,2$,
with polynomial growth such that, for $i=1,2$ and any $(t,x)\in
[0,T]\times \R^{k}$,
\begin{equation}\label{eq:Feynmankacrep_viscositysol}
(Y_{s}^{1,(t,x)},Y_{s}^{2,(t,x)})=(u^1(s,X^{t,x}_s),u^2(s,X^{t,x}_s)),\,\,\forall
s \in [t,T].
\end{equation}

Now, and in the same fashion, when considering the decreasing approximating scheme, it follows from the
same result in \cite{Elkarouietal97} and from the uniqueness that $u_1$
and $u_2$ are also upper semi-continuous. Therefore, $u_1$ and $u_2$ are continuous with polynomial growth. Finally, relying on Theorem 8.5
in  \cite{Elkarouietal97} we directly obtain the following result.
\begin{thm} The pair $(u^1,u^2)$ defined in (\ref{eq:Feynmankacrep_viscositysol}) is a continuous viscosity solution for the system
$(VI)$. 
\end{thm}
\begin{remark} The question of uniqueness of the system $(VI)$ in the general case is more involved and will appear elsewhere.
Actually, in using the results of e.g. the second example of Section 3.2, we easily see that uniqueness for $(VI)$ does not hold in general.
\end{remark}

\end{document}